\documentclass{epl}

\usepackage{psfig}
\title{Chemical fracture statistics and universal distribution of
extreme values.}
\author{A. Baldassarri\inst{1-3} \and A. Gabrielli\inst{1,3} \and B. 
Sapoval\inst{1,2}
}
\shortauthor{A. Baldassarri \etal}
\shorttitle{Chemical fracture and distribution of extreme values}
\institute{
  \inst{1} Lab. de Phys. de la Mati\`ere 
Condens\'ee, Ecole Polytechnique 91128 Palaiseau, France \\
  \inst{2} CMLA, 
Ecole Normale Sup\'erieure, 
94235 Cachan, France\\
\inst{3} INFM - Dipartimento di Fisica, Univ. di Roma ``La 
Sapienza'', 
P.le A. Moro, 2, ls I-00185 Roma, Italy
}
\begin{document}

\maketitle

\begin{abstract}
When a corrosive solution reaches the limits of a solid sample, a
chemical fracture occurs. An analytical theory for the probability of
this chemical fracture is proposed and confirmed by extensive
numerical experiments on a two dimensional model. This theory follows
from the general probability theory of extreme events given by Gumbel.
The analytic law differs from the Weibull law commonly used to
describe mechanical failures for brittle materials.
However a three parameters fit with the Weibull law gives good
results, confirming the empirical value of this kind of analysis.
\end{abstract}

Chemical etching of disordered solids is an important technological
problem \cite{corr-exp}, that presents as well interesting questions
in the theory of random systems\cite{herrmann}.  Strong etching
solutions will eventually lead to the fracture of a finite sample, an
event defined as ``chemical fracture''.  In this letter we present a
theory for the statistical behavior of this specific fracture
mechanism.  We show \cite{these} that the deep nature of the {\it
chemical} fracture statistics relies on the probabilistic theory of
extreme events due to Gumbel~\cite{Gumbel}. Interestingly it is found
that our {\em chemical} fracture statistics are not practically
distinguishable from the Weibull statistics empirically introduced to
fit {\em mechanical} fracture statistics \cite{Weibull}.

The problem is studied using a simple two-dimensional corrosion
picture \cite{sapo-etch} inspired by an experimental study of pit
corrosion of aluminum films \cite{balazs}.  The model describes the
chemical etching of a random solid by a finite volume of solution. It
predicts that the etching process stops spontaneously on a fractal
liquid-solid interface as observed experimentally.

\begin{figure}
\onefigure[angle=-90,width=7cm]{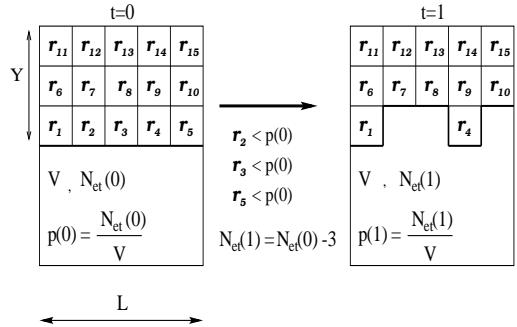}
\caption{Sketch of the etching dynamics in a square lattice: the sites
$2,3,5$ are etched at the first time-step as their resistances are
lower than $p(0)$.  At the same time the number of etchant particles
in the solution decreases by $3$ units, and a new part of the solid is
uncovered. The process is then iterated.}
\label{model}
\end{figure}

The two dimensional system is described in Fig.~\ref{model}: The solid
is made of lattice sites exhibiting random ``resistances to
corrosion'' $r_{i}\in [0,1]$ uniformly distributed.  It has a width
$L$ and a depth $Y$. At any time $t$ the ``etching power'' of the
solution is proportional to the etchant concentration $C(t)$ :
$p(t)=\Gamma C(t)$. The solution has a volume $V$ and contains an
initial number $N_{et}(0)$ of etchant molecules.  Therefore
$p(0)=C(0)=N_{et}(0)/V$ using $\Gamma=1$. Hereafter we choose $p(0)>
p_c$, where $p_c$ is the percolation threshold of the lattice.

The solution is initially in contact with the solid through the bottom
boundary $y=0$.  At each time-step $t$, all surface sites with
$r_{i}<p(t)$ are dissolved and a particle of etchant is consumed for
each corroded site.  Hence, the solution concentration progressively
decreases.  For an arbitrarily large sample, the corrosion process
always stops spontaneously at a certain time $t_f$ when the etchant
concentration $p(t_f)$ is still finite \cite{pap-gen}, reaching a
maximal depth $y_{M}$.  However, when $Y$ is finite, the corrosion can
reach the bottom of the sample before $t_f$, fracturing the solid into
two disconnected parts.  This "chemical fracture" is a stochastic
event which depends on the realization of the random resistances
$r_i$.

\begin{figure}
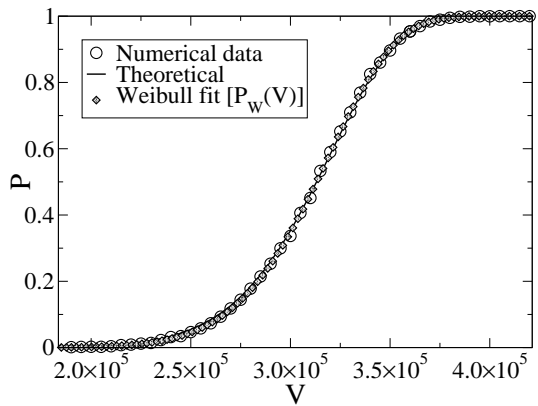

\onefigure[width=7cm]{figure2.eps}
\caption{{\em Chemical fracture} probability $P\left(y_M>Y\right)$ as
a function of the volume $V$ which measures the ``chemical force''.
Small circles are the numerical estimated probabilities performing
$1000$ runs for each value of $V$ applied on a solid of sizes $L=1000$
and $Y=500$ ($p_0=0.7$).  Diamonds represent the fit of the fracture
data with a Weibull law of parameters $V_0= 1.15\cdot 10^5$,
$V_1=4.08\cdot 10^5$, and $m=3.89$.  The line is the result of the fit
using Eq.~(\protect{\ref{gum-wei-s}}), as explained in the text (the
fitted parameter $A= 0.3875$).  }
\label{gum-vs-wei}
\end{figure}

Hereafter we call ${\cal P}(V)$  the fraction of samples broken by a
volume $V$ of solution, keeping the other system parameters $L$, $p_0$
and $Y$ fixed. Its measure is shown in Fig.~\ref{gum-vs-wei}. The
solution volume represents the applied chemical force for given
$p_0$. Obviously the fracture event is directly related to the value
of the maximal depth $y_M$ relative to the sample depth $Y$: if
$y_M<Y$ then the fracture does not occur, otherwise the sample will be
broken in two distinct pieces. Hence, in order to compute the fracture
statistics we have to study the probability distribution of $y_M$ in a
sample with an infinite depth.

Before proceeding, it is important to remark the differences between
the {\em chemical fracture statistics} and the {\em distribution of
the maximal corrosion depth}.  The {\em chemical fracture statistics}
${\cal P}(V)$ is a {\em parametric} curve which gives the fraction of
similar solid samples fractured for a given value of $V$. Hence ${\cal
P}(V)$ is not a distribution function in the sense of probability
theory. In particular $V$ is not a random variable but an external
parameter.

On the contrary, the {\em distribution of the maximal corrosion depth}
is the probability distribution function ({\bf DF}) of the random
variable $y_M$, representing the maximal corrosion depth for an
(infinitely deep) sample. This is a genuine probability distribution
function and, accordingly, its derivative is the probability density
function ({\bf pdf})  of the variable $y_M$. Obviously the two
quantities are related, but of a different nature. In the following we
will show how it is possible to recover the first, via a probabilistic
theory of the second.

The fracture statistics ${\cal P}(V)$ of Fig. 2 has been measured
numerically taking a sample of a fixed depth and performing several
corrosion processes for each value of the solution volume $V$, and
recording the fraction of broken samples as a function of $V$ (see
Fig.~\ref{gum-vs-wei}).

\begin{figure}
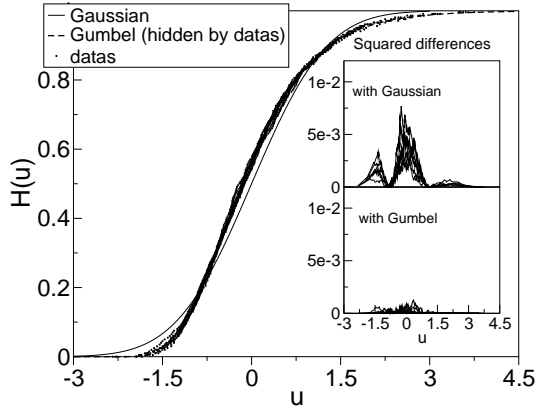

\onefigure[width=7cm]{figure3.eps}
\caption{The numerical integrated probability distribution for reduced
$y_M$ (zero mean and unitary variance) is compared with standard
Gaussian and Gumbel distributions. In the main figure, the simulation
results collapse on the Gumbel distribution given by
Eq.~(\ref{gumb-gen}) (hidden by data points).  In the insets we show
the squared difference between data and the Gaussian and Gumbel
distributions. The data refer to simulations performed for
$L=3000,5000$ and $N_{et}(0)=5\cdot 10^6, 1\cdot 10^7, 2\cdot 10^7,
5\cdot 10^7$ ($1000$ different dynamics for each choice of $L$ and
$V$, $p_0=0.7$).  }
\label{ymax-distr}
\end{figure}

On the other hand, the probability distribution of the maximal
corrosion depth have been measured fixing $V$ and recording the
maximal corrosion depth during several corrosion processes on large
systems. A different choice of $V$ give a different distribution, but
a universal scaling form can be recognized.  As shown in
Fig.~\ref{ymax-distr} the numerical DF $P(z_M<u)$ for the reduced
variable $z_M\equiv (y_M-\left<y_M\right>)/\Sigma$ (where
$\left<y_M\right>$ is the average value of $y_M$ and $\Sigma$ its
standard deviation) is the same for several choices of $V$.

Since $y_M$ is an extremal random variable, we compare this numerical
distribution with the {\em standard} (i.e. with zero mean and unitary
variance) Gumbel distribution function $H(u)$ \cite{Gumbel} for
extreme events:
\begin{equation}
H(u)=e^{-e^{-(bu+a)}}\,,
\label{gumb-gen}
\end{equation}
where $a\simeq -0.5772$ and $b\simeq \sqrt{1.64493}$. 

We recall that the Gumbel distribution is the canonical distribution
for the maximal value of a set of random variables in the same manner
that the Gaussian distribution is the canonical distribution for the
sum. More precisely if $\{x_1,...,x_N\}$ is a set of independent
stochastic variables with identical distribution decaying faster than
any power law for large values, then, in the large $N$ limit, the
maximum value among them can be shown to be Gumbel distributed.  The
collapse of our data on the Gumbel distribution is remarkable, first
because there are no adjustable parameters, and second because the
points on the interface {\it are correlated to some extent}.

In the following we discuss why the physics of the etching model
determines such a Gumbel behavior.  As a second step we will derive
from it the theoretical form of ${\cal P}(V)$.  This last task is
accomplished considering the probability $P(y_M>Y)$ and introducing
the explicit dependence of $\left<y_M\right>$ and $\Sigma$ on $V$
(with $p_0$, $Y$ and $L$ fixed), i.e.:
\begin{equation}
{\cal P}(V)\!
=\!1-\exp\!\left\{-\exp\!\left[b\frac{\left<y_M\right>\!(V)-Y}{\Sigma(V)}-
a\right]\right\}
\label{gum-wei-s-bis}
\end{equation}
Such explicit dependence is obtained using
the known scaling results of this 
particular etching model.

\begin{figure}
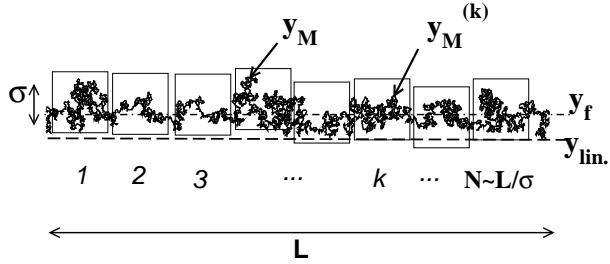

\onefigure[width=8cm]{figure4.eps}
\caption{Final corrosion front: It is composed of $N\sim L/\sigma $
independent regions of size $\sigma$ ($\sigma$ can be considered as
the correlation length of the system). The extremal front position
$y_M$ is indicated. It is the max between the $N$ independent values
of $y_M^{(k)}$ of each region.  }
\label{interface}
\end{figure}

We now proceed with the derivation of the Gumbel law for this etching
model. First we have to explain why the corrosion front can be reduced
to a set of independent random variables.  At the beginning of the
dynamics \cite{pap-gen}, $p(t)$ follows the exponential law $p(t) =
p_0\exp (-t/\tau)$ with $\tau=V/L$. The corrosion front advances layer
by layer up to approximatively $t=t_c$ when $p(t)=p_c$ and reaches a
depth $y_{lin.}\sim V/L$. After this period the corrosion front
becomes very irregular and finally stops at $t=t_f$.  At $t_f$ the
etching power $p_f=p(t_f)$ is slightly smaller than $p_c$ and the
final corrosion front is fractal with dimension $D_{f}=7/4$ up to a
characteristic width $\sigma< L$. This situation is displayed in
Fig.~\ref{interface}.  As shown in~\cite{pap-gen}, this phenomenon
obeys the scaling laws of Gradient Percolation~\cite{gp} where the
role of the gradient is played by the ratio $L/V$.  This implies that
\begin{itemize}
\item $\sigma \sim (L/V)^{-1/D_f}$,
\item $\sigma$ can be seen as a percolation
correlation length.
\end{itemize}
The total final corrosion front, shown in Fig.~\ref{interface}, can
then be considered as a juxtaposition of $N\sim L/\sigma$ nearly
independent fractal boxes of lateral width $\sim\!\sigma$ (for a similar recognition
of effective independent variables in extreme statistics see \cite{curtin}).  Inside
each box $k$, there is a point of maximal penetration of the front
$y_M^{(k)}$.  Since such points belong to different boxes, the values
$y_M^{(k)}$ are, by construction, identically distributed and
independent random variables.  The extreme position of the front is
the maximal value between such a collection of random variables.  This
is why a Gumbel distribution is observed.

More precisely due to the underlying percolation phenomena, the DF for
each $y_M^{(k)}$ has exponential tail with the same characteristic
scale $\sim\!\sigma$:
\begin{equation}
P\left( y_M^{(k)}>y\right) \sim e^{-\frac{y-\left\langle 
y_M^{(k)}
\right\rangle}{c\sigma }}
\;\;\mbox{for}
\;y-\left\langle y_M^{(k)}\right\rangle> \sigma \,,  
\label{y-tail}
\end{equation}
where $c$ is a constant of order 1 (see also some recent results on
the size of sub-critical clusters~\cite{bazant}).  In this case, the
theory of extreme statistics\cite{Gumbel} imposes that, for $N\gg1$,
$y_M$ is asymptotically Gumbel distributed, i.e.
$P\left[(y_M-\left<y_M\right>)/\Sigma)<u\right]=H(u)$ and that:
\begin{eqnarray}
\left\langle y_M \right\rangle & = & 
\left\langle y_M^{(k)}\right\rangle+ 
\Sigma \log N 
\label{mom1-y}\\  
\Sigma= & 2c\sigma\;.
\label{mom2-y}
\end{eqnarray}
It is then necessary to test numerically Eqs.~(\ref{mom1-y})
and~(\ref{mom2-y}) (and specifically the logarithmic dependence on
$N$, i.e. on $L/\sigma$), in order to confirm this arguments.  In the
situation described in Fig.~\ref{interface} one observes that
$\left\langle y_M^{(k)}\right\rangle $ is given by the average depth
$y_f$ of the final corrosion front (average taken over all the final
front sites) plus a positive shift of order $\sigma$.  In turn, the
average depth $y_f$ (a directly measurable quantity in simulations) is
given by the depth $y_{lin.}$ reached during the linear part of the
corrosion in addition to a shift (sub-leading) again of order
$\sigma$.  Therefore from Eq.~(\ref{mom1-y}) we can write:
\begin{eqnarray}
\langle y_M\rangle-y_f&\sim& 
\sigma\left[1+k_1\ln\left(L/\sigma\right)\right]\label{scaling1}\\
y_f-y_{lin.}&\sim&\sigma\label{scaling2}\\
y_{lin.}&\sim& V/L\label{scaling4}\\
\Sigma&\sim&\sigma\label{scaling3}
\end{eqnarray}
where $k_1$ is a coefficient of order one.  Recalling that
$\sigma\sim(V/L)^{4/7}$, we can substitute $\left<y_M\right>$ and
$\Sigma$ as functions of $V$ in Eqs.~(\ref{mom2-y}) and (\ref{mom1-y})
in order to obtain the {\em real} functional dependence of ${\cal
P}(V)$ on $V$ apart from numerical scaling coefficients.

\begin{figure}
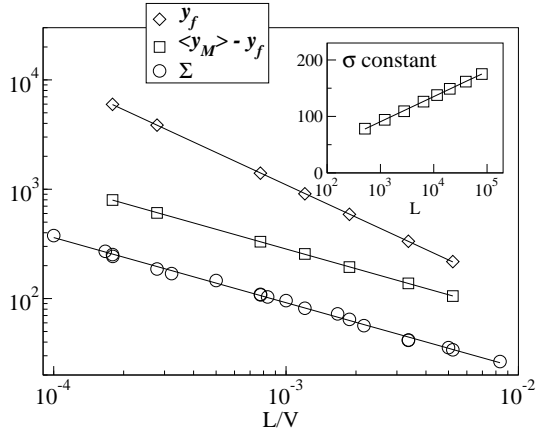

\onefigure[width=7cm]{figure5.eps}
\caption{ Scaling behavior of $y_f$, $\left[\left\langle
y_{M}\right\rangle-y_f\right]$ and $\Sigma$.  The data in the main
graph refer to simulations performed with a constant $L/\sigma$ (range
of parameters: from $L=522$ and $V=5\cdot 10^8$ to $L=11550$ and
$V=5\cdot 10^8$).  The numerical fits are in accordance with the
theoretical behaviors.  The inset shows the logarithmic dependence of
$\left<y_M\right>$ on $L$ for a given $\sigma$ (parameters range from
$L=1205$ and $V=120500$ to $L=80000$ and $V=8\cdot 10^6$, chosen to
have a constant $\sigma$).  In both cases the data correspond to
$1000$ different dynamics for each choice of $L$ and $V$, $p_0=0.7$.
}
\label{fig-yf}
\end{figure}

The scaling behaviors are themselves confirmed by the extensive
numerical simulations shown in Fig.~\ref{fig-yf}. Note that if
$L/\sigma$ is kept constant (i.e. if $L\sim V^{4/11}$) then
$[\left\langle y_M\right\rangle-y_f] \sim (L/V)^{-1/D_f}$.  The
measured exponent $\ 0.59 \pm 0.02$ is consistent with this
prediction.  In the same figure, we also report the scaling behavior
of the standard deviation $\Sigma$ and of $y_f$.  These fits confirm
that $\Sigma\sim \sigma$ and that $y_f$ can be written as the sum of
$y_{lin.}$ and a shift of order $\sigma$.  Finally, the linear
logarithmic dependence on $L$ for fixed $\sigma$ (i.e. for fixed
$L/V$), which is the essential confirmation of Eq.~(\ref{mom1-y}), is
shown in the inset.  It has also been numerically confirmed that
$\Sigma$ and $y_f$ does not depend on $L$ for fixed sigma (not shown
in the figure).

To summarize, Eq.~(\ref{gum-wei-s-bis}) and Eqs.~(\ref{scaling1}-
\ref{scaling3}) give the theoretical form for ${\cal P}(V)$, once the
proportionality coefficients in Eqs.~(\ref{scaling1}-\ref{scaling3})
and the coefficient $k_1$ have been determined.  For simplicity, it is
convenient to rewrite Eq.~(\ref{gum-wei-s-bis}) as follows:
\begin{equation}
{\cal P}(V)\!=\!1\!-\!\exp \left\{ -A\left(\frac{L}
{\Sigma}\right)^b\!\!\exp \left[ \frac{b(y_f
-Y)}{\Sigma} \right] \right\} \,.
\label{gum-wei-s}
\end{equation} 
The dependence on $k_1$, described in Eq.~(\ref{scaling1}), is now
contained in the coefficient $A$.  As a closure test of our analysis
we choose to use this form to fit against $A$ the numerical fracture
statistics (i.e. the fraction of simulation runs breaking the sample
as a function of $V$)~\cite{A}.  As shown in Fig.~\ref{gum-vs-wei},
the theoretical form of ${\cal P}(V)$ and the simulation are in
excellent agreement.

In summary, we have studied the chemical fracture due to the extremal
propagation of a corrosion front in a two-dimensional scalar etching
model. It has been shown both theoretically and numerically that the
statistics of the maximal depth reached by the solution is given by
the well known Gumbel extremal distribution function. From this it has
been possible to extract theoretically the probability of a {\em
chemical fracture}.

Before concluding, we believe that it could be interesting to compare
our results with the commonly analysis used for mechanical fracture
statistics. In this case, even though a widely accepted theoretical
frame is still lacking, the standard empirical fit is often performed
using a Weibull law \cite{Weibull}.  In our chemical fracture model,
this comes to try to fit the datas with:
\begin{equation}
{\cal P}_W(V)=1-e^{-\left( \frac{V -
V_{0}}{V_{1}}\right)^{m}}\;\;\;\;\mbox{with}\;V>V_0\,,
\label{weibul}
\end{equation}
where $V_0$ (the minimal stress to have a finite fraction of
fractures), $V_1$ and $m$ are suitable parameters.

The result, shown in Fig. 2, is that the chemical fracture statistics can
be fitted very accurately with a Weibull law.  We have also verified (not
shown here) that the Weibull low probability tails (the test generally
used by engineers) fits nicely with our data. On one side, we believe that
this is astonishing, since, despite the simple scalar nature of the model,
we recover, for our purely chemical
process, the same empirical statistics found for mechanical failures of
brittle materials, generally observed in more complex (vectorial or
tensorial) frames. On the other side, it is important to remark that in
the present case the Weibull law is only an empirical fit, because the
real analytical form of the chemical fracture statistics is given by
Eq.\ref{gum-wei-s-bis}. Moreover a Weibull law, which is usually related
to the presence of power law distributed flaws, has no reason to emerge
here, as Eq.\ref{gum-wei-s-bis} results from a theoretical
calculation of the model and directly connected to known results in
percolation theory.

As a final remark, one could suggest that the correspondence between
the Weibull fit and the underlying Gumbel distribution (first
suggested by \cite{duxbury}) found here is linked more generally to
Gradient Percolation.  Gradient Percolation situations can also be
present in solids which {\it have never been etched}~\cite{zap}.  The
essential ingredients of our theory are (i) the percolation aspect
linked to randomness of the system, and (ii) the existence of a
gradient.  In real systems gradients may exist in the sample, for
example the ion concentration gradient in glass fibers used in
communication optics. But it may also appear in experiments, for
instance in experimental studies of ceramic fracture. In classical
flexure experiments the distribution of stress is non uniform
\cite{Jaya}. It remains to be determined specifically if such gradient
mechanisms are responsible for some of the ubiquitous apparent Weibull
statistics. 

Our two dimensional results could be generalized to higher dimensions,
according to~\cite{field}. Therein, field theoretic arguments show that
the corrosion dynamics towards the final state can be described as a
self-organized absorbing phase transition towards the critical phase
of percolation in any dimension.

\acknowledgments
We are grateful to M.~Dejmek and S.~Zapperi for useful comments and we 
acknowledge the support of the European Community TMR Network 
ERBFMRXCT980183.


\begin{thebibliography}{00}
\bibitem{A} For  $\Sigma$ and $y_f$
we impose the scaling predictions
Eq.~(\protect{\ref{scaling1}}-\protect{\ref{scaling3}}) 
with coefficients independently 
fitted from data in Fig.~\ref{fig-yf}. From the same data 
it should be possible to get $k_1$, and hence $A$. 
However, the logarithmic nature of the scaling makes this procedure
delicate, and we prefer to fit $A$ instead.
\end{thebibliography}
\end{document}